\documentclass[useAMS,usenatbib]{mn2e}

\usepackage{epsfig}
\title[Lens Inversion of RXS J1131-1231]{Unlensing HST Observations of the Einstein Ring 1RXS J1131-1231: A Bayesian Analysis}
\author[B. J. Brewer and G. F. Lewis]{B. J. Brewer\thanks{E-mail:
brewer@physics.usyd.edu.au} and G. F.
Lewis\\
Institute of Astronomy, School of Physics, The University of Sydney, NSW, 2006, Australia}

\begin{document}

\date{\today}

%\pagerange{\pageref{firstpage}--\pageref{lastpage}} \pubyear{2007}

\maketitle

\label{firstpage}

\begin{abstract}
We present a source and lens reconstruction for the optical Einstein ring gravitational lens system RXS J1131-1231. We resolve detail in the source, which is the host galaxy of a $z=0.658$ quasar, down to a resolution of 0.045 arc seconds (this is the size of the smallest conclusively resolved structures, rather than the pixel scale), using a Bayesian technique with a realistic model for the prior information. The source reconstruction reveals a substantial amount of complex structure in the host galaxy, which is $\sim$ 8 kpc in extent and contains several bright compact substructures, with the quasar source residing in one of these bright substructures. Additionally, we recover the mass distribution of the lensing galaxy, assuming a simply-parameterised model, using information from both the quasar images and the extended images. This allows a direct comparison of the amount of information about the lens that is provided by the quasar images in comparison to the extended images. In this system, we find that the extended images provide significantly more information about the lens than the quasar images alone, especially if we do not include prior constraints on the central position of the lens.
\end{abstract}
\section{Introduction}
Gravitational lensing can be used as a powerful astrophysical tool for probing the density profiles of galaxies, and is one of the few ways in which dark matter can be detected \citep[e.g.][]{2005MNRAS.363.1136K}. In addition, it often magnifies source objects by one to two orders of magnitude. This allows us to use the intervening gravitational lens as a kind of natural telescope, magnifying the source so that we can observe more detail than we would have been able to without the lens. This extra magnification provided by lensing has been very beneficial to studies of star formation and galaxy morphology at high redshift. Regions of the galaxy size and luminosity distribution that are inaccessible in unlensed observations are made (more) visible by lensing \citep[e.g.][]{2000ApJ...528...96P, wayth, 2006ApJ...651....8B, 2007ApJ...671.1196M, 2008arXiv0804.4002D}. The properties of the lens galaxies (typically elliptical galaxies) can also be inferred from their lensing effect \citep[e.g.][]{2006ApJ...649..599K, 2008arXiv0806.1056T}. Of course, gravitational lensing distorts the image of the source, as well as magnifying it. Thus, techniques have been developed that aim to infer the mass profile of the lens galaxy and the surface brightness profile of the source, given observed images \citep[e.g.][]{2003ApJ...590..673W, 2006ApJ...637..608B}.

The aim of this paper is to carry out this process with the recently discovered gravitationally lensed quasar/host galaxy system RXS J1131-1231 \citep{2003A&A...406L..43S}. This system consists of a quadruply imaged quasar at redshift $z=0.658$ lensed by a galaxy at $z=0.295$. At the time of its discovery, it was the closest known lensed quasar, with some evidence for an extended optical Einstein ring - the image of the quasar host galaxy. Initial simple modelling suggested that the quasar source was magnified by a factor of $\sim$ 50. Thus, subsequent observations with the ACS aboard the Hubble Space Telescope \citep[][hereafter C06]{2006A&A...451..865C} allow the recovery of the  morphology of the quasar's host galaxy down to a resolution of about 0.01 arc seconds - at least in principle, for the parts of the source nearest the caustic. Indeed, C06 presented a wide array of results based on HST observations (at 555nm and 814nm with ACS, and 1600nm with NICMOS), including a detailed reconstruction of the extended source.

The source reconstruction method used by C06 is based on lensing the image back to a pixellated grid in the source plane, setting the source surface brightnesses to equal the image surface brightness, and using a decision rule (in this case, the median) to decide on the value of a source pixel whenever two or more image pixels disagree about the value of the same source pixel. If the point spread function (PSF) is small or the image has been deconvolved (in C06, the deconvolution step was neglected for the purposes of the extended source reconstruction) and the lens model is correct, this method can expect to be quite accurate. However, in principle, the uncertainty in the lens model parameters and the deconvolution step should always be taken into account. In this paper, we focus our attention on the 555nm ACS image (the drizzled images, as reduced by C06, were provided to us), and the process by which we reconstruct the original, unlensed source from it. Any differences between our reconstruction and the C06 one can be attributed to the following advantages of our approach: PSF deconvolution, averaging over the lens parameter uncertainties, simultaneous fitting of all parameters, and the prior information that Bayesian methods are capable of taking into account: in the case of our model, that is the knowledge that the majority of pixels in an astrophysical sources should be dark \citep{2006ApJ...637..608B}. The 555nm image is also of particular interest because its rest wavelength (335nm) probes regions of recent star formation in a galaxy with an AGN.

In the case of the Einstein Ring 0047-2808 \citep{2006ApJ...651....8B}, our method was able to resolve structure down to scales of $\sim$ 0.01 arcsec, a factor of five smaller than that obtainable in an unlensed observation with the Hubble Space Telescope and about double the resolution obtained by \citet{2005ApJ...623...31D} using adaptive pixellation and least squares {\it applied to exactly the same data}. This was possible because we used a prior distribution over possible sources that is more realistic as a model of our knowledge of an unknown astrophysical source, that is, we took into account the fact that it should be a positive structure against a dark background, a fact many methods (such as least squares and some popular regularisation formulas) implicitly ignore \citep{2006ApJ...637..608B}. These differences between methods are likely to be most significant when data are sparse or noisy, and all methods tend to agree as the data quality increases and we approach the regime where the observed image uniquely determines the correct reconstruction.
\section{Background to the Method}
The conceptual basis of the Bayesian reconstruction method was presented in \citet{2006ApJ...637..608B}. The idea is to fit a complex model to some data, but rather than simply optimising the parameters of the model to achieve the best fit, we try to explore the whole volume of the parameter space that remains viable after taking into account the data. The effect of data is usually to reduce the volume of the plausible regions of the parameter space considerably\footnote{For non-uniform probability distributions, ``volume'' is effectively the exponential of the information theory entropy of the distribution.}. The exploration of the parameter space can be achieved by using Markov Chain Monte Carlo (MCMC) algorithms, which are designed to produce a set of models sampled from the posterior distribution. In the case of modelling the background source and lens mass distribution of a gravitational lensing system, this allows us to obtain a sample of model sources and lenses that, when lensed and blurred by a PSF, match the observational data. The diversity of the models gives the uncertainties in any quantity of interest. The reader is referred to \citet{gregory} for an introduction to Bayesian Inference and MCMC.
\section{Method and Assumptions}
The first step of a Bayesian analysis is to assign a likelihood function, or the probability density we would assign to the observed data if we knew the values of all of the parameters. To assign this, we need a noise frame, a measure of how uncertain we are about the noise level in each pixel. This is typically done by assuming that the observational error at pixel $i$ is from a normal distribution with mean 0 and known standard deviation $\sigma_i$. We extended this to include two ``extra noise parameters'' $\sigma_a$ and $\sigma_b$, such that the standard deviation for the error in the $i$th pixel is $\sqrt{\sigma_i^2 + \sigma_a^2 + \sigma_b^2 f_i}$, where $f_i$ is the predicted flux in the $i$'th pixel. $\sigma_a$ and $\sigma_b$ then become extra model parameters to be estimated from the data. The inclusion of $\sigma_a$ and $\sigma_b$ implies that the extra noise level sigma varies with the predicted brightness of the pixel, with a square root dependence expected from Poisson photon counting.

\begin{figure*}
\begin{center}
\includegraphics[scale=0.6]{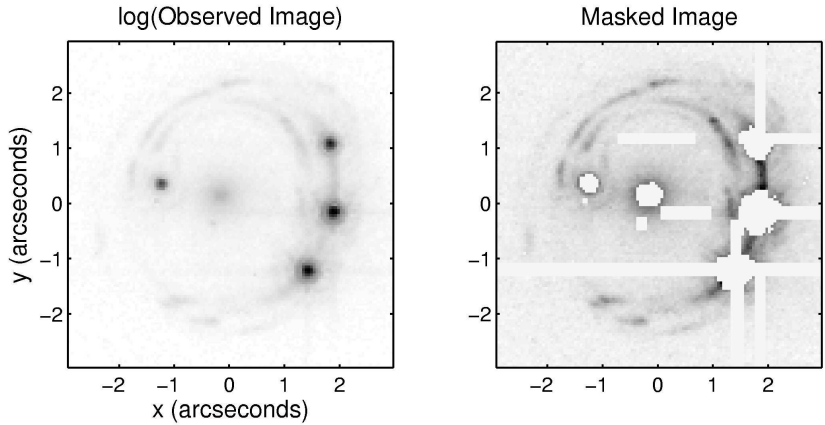}
\end{center}
\caption{The logarithm of the observed image (on the left) shows both the quasar images, along with their additional effects such as diffraction spikes, and the faint Einstein ring image of the host galaxy. On the right is the image (scaled linearly) with some parts blocked out - these blanked regions are those where the $\{\sigma_i\}$ have been adjusted to block out the brightest emission from the quasar. The $\sigma_i$ for some pixels has also been artificially boosted to reduce the effects of the outer parts of the diffraction spikes from the quasar images. We expect the inner parts of these spikes to be well modelled by the TinyTim profiles (Figure~\ref{tinytim}). Note that the lens galaxy light profile extends over this entire image; where it appears that the image is blank, there is actually a positive flux.\label{data}}
\end{figure*}

We chose the $\{\sigma_i\}$ values to be zero for most of the image, but infinite for the brightest regions of the quasar images, effectively masking out those parts of the image; this mask can be seen in Figure~\ref{data}.

A model PSF was obtained using the TinyTim software \citep{tinytimreference}. As noted by C06, the TinyTim simulations did not successfully perform the geometric distortion correction, and the output PSF had slightly non-orthogonal diffraction spikes, whereas the spikes in the image are perpendicular. To correct this, the PSF was ``straightened out'' by evaluating it with respect to non-orthogonal axes; the resulting PSF is shown in Figure~\ref{tinytim}. Whilst this process is imperfect, the extra noise sigma protects against serious consequences resulting from slight inaccuracies in this process. While our choice of $\{\sigma_i\}$ was designed to block out the brightest parts of the quasar images, since they are so bright, light from the quasar images still extends out past the masked regions and overlaps with interesting Einstein Ring structures. Thus, when modelling the image, we still required a flux component due to the quasar images.

The four quasar images were modelled as being proportional to the corrected TinyTim profiles with unknown fluxes and central positions. The surface brightness profile of the lens galaxy was modelled as the sum of two elliptical Gaussian-like profiles (Sersic profiles, one for the core and one for the extended emission) proportional to $e^{-(\frac{R}{L})^{\alpha}}$, where $R = \sqrt{Qx'^2 + y'^2/Q}$ with unknown peak surface brightness, ellipticity $Q$, length scale $L$, and angle of orientation. The central position was also considered initially unknown, but for MCMC purposes the starting point was to have both profiles centred near the observed centre of the lens galaxy core. The slopes $\alpha$ were restricted to the range $[0,10]$ and assigned a uniform prior distribution, along with all of the other free parameters. Although elliptical galaxies are well modelled by a Sersic profile with $\alpha = 1/4$, we are modelling this galaxy by {\it two} such profiles. This was done because the wings of the lens galaxy light profile (where it overlaps the Einstein ring) are of great significance for our source reconstruction, and we do not want the core of the lens galaxy to be relevant to the wings.

Note that there are parts of the image where flux is present from three sources: the lensed Einstein Ring, the wings of the PSF from the quasar images, and the foreground lens galaxy. The fact these all overlap suggests that the optimal approach (in all senses apart from CPU time) involves simultaneously fitting all of these components. Throughout this paper, any modelling has included all of the lens galaxy profile parameters as free parameters\footnote{Except for the first part of Section~\ref{pixellated}, where computational restrictions required that we fix the lens parameters.}, as well as the source, lens model parameters and positions and brightnesses of the four TinyTim PSF profiles, to model the contribution from the quasar images that remains even after masking out their central regions.
\section{Lens Model Parameterisation}
The particular lens model we assumed for this system was a pseudo-isothermal elliptical potential (PIEP) \citep{piepref}, primarily for computational speed but also because it is fairly general and realistic, at least for single galaxy lenses that are not too elliptical. This model has five parameters: strength $b$, ellipticity $q$ (actually the axis ratio: $q=1$ implies a circularly symmetric lens), orientation angle $\theta$, and two parameters $(x_c,y_c)$ for the central position (measured relative to the central pixel of the images in Figure~\ref{data}). Although any Bayesian modelling can only explore a particular slice through the full hypothesis space we might have in our minds, using a simplified analytical model is often sufficient to illuminate the general properties of the true lens mass distribution. Also, it is typically the case that inferences about the source of an Einstein Ring are insensitive to the specific parameterisation for the lens model \citep[e.g.][]{wayth}, as long as the model is able to fit the observed image at all.

All Einstein rings can be expected to reside in an environment where the external shear due to neighbouring galaxies is nonzero (Kochanek, private communication), and thus, external shear was also included in the lens model. There are two parameters for the external shear: $\gamma$, its magnitude, and its orientation angle $\theta_{\gamma}$. \citet{2007ApJ...660.1016S} have observed the flux ratios of the quasar images (via integral field spectroscopy) and found that most of these ratios are consistent with a model of this type (elliptical potential plus external shear).  A similar model was used by C06 (they used a singular isothermal ellipse+$\gamma$), where they find that it is the simplest parameterised model that can reproduce the observations. In principle, we could adopt ever less restrictive parameterisations for the lens, to hunt for substructures in its density profile. However, such an approach is extraordinarily computationally expensive (unless simplifying assumptions about the source are also made) and is beyond the scope of this paper.
In terms of all of these parameters, the deflection angle formula, relating a point $(x,y)$ in the image plane to a corresponding point $(x_s,y_s)$ in the source plane, is
\begin{eqnarray}\label{lenseqn}
x_s = x - \alpha_x(x,y) \nonumber \\
y_s = y - \alpha_y(x,y)
\end{eqnarray}
where the deflection angles $\alpha$ are given by the gradient of the potential
\begin{equation}
\psi(x,y) =  \frac{1}{2}\gamma (x_{\gamma}^2 - y_{\gamma}^2) + b\sqrt{qx_{\theta}^2 + y_{\theta}^2/q}
\end{equation}
and $(x_{\gamma},y_{\gamma})$ are the ray coordinates in the rotated coordinate system whose origin is $(0,0)$ and is aligned with the external shear (i.e. rotated by an angle $\theta_{\gamma}$), and $(x_{\theta},y_{\theta})$ are the ray coordinates in the rotated and translated coordinate system centred at $(x_c,y_c)$ and oriented at an angle $\theta$. The physical interpretation of each of these parameters suggests a plausible prior range for their values. To represent this knowledge we used the following prior distributions (Table~\ref{lenspriors}). Since these are broad distributions, and the data are good, the influence of these choices is negligible; they are included for completeness.
\begin{table*}
\begin{center}
\caption{Prior  probability densities for  the lens  model parameters,
and also the  extra noise parameters $\sigma_a$ and $\sigma_b$.}\label{lenspriors}
\begin{tabular}{lcc}
\hline Parameter & Prior Distribution  \\
\hline $b$ & Normal, mean 1.8'', SD 0.5, $b > 1$\\
$q$ & Normal, mean 0.9, SD 0.2, $0 < q < 1$\\
$x_c$ & Normal, centred at the lens galaxy core, SD 1.0'' \\
$y_c$ & Normal, centred at the lens galaxy core, SD 1.0'' \\
$\theta$ & Uniform, between 0 and $\pi$ \\
$\sigma_a$ & Improper Uniform, $\sigma_a > 0$ \\
$\sigma_b$ & Improper Uniform, $\sigma_b > 0$ \\
$\theta_{\gamma}$ & Uniform, between 0 and $2\pi$ \\
$\gamma	$ & Exponential, mean 0.1
\end{tabular}
\medskip\\
\end{center}
\end{table*}

\begin{figure}
\begin{center}
\includegraphics[scale=0.4]{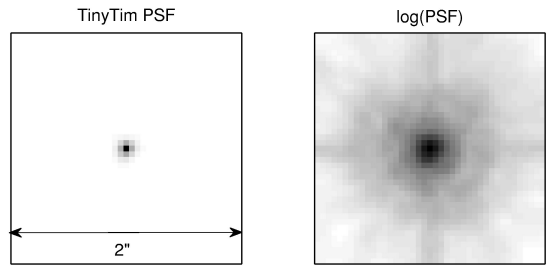}
\end{center}
\caption{Simulated PSFs from TinyTim. Each pixel corresponds to an ACS image pixel and is 0.049 arcseconds across. On the left is the actual PSF. For the purposes of the modelling of the extended images, this PSF was truncated to a 5$\times$5 pixel PSF using only the brightest central parts of the PSF. For the quasar images, the wings are most important, and this can be seen most easily in the right hand panel. Since the quasars are so bright, the wings of the PSF are not negligible.\label{tinytim}}
\end{figure}
In summary, the observed image was modelled as the sum of the following components:\\
(i)	Four TinyTim PSF profiles (Figure~\ref{tinytim}) with initially unknown amplitude and central position, to model the quasar images. While the bright parts of the images are masked out, the wings of the PSF are still important, so these components are required.\\
(ii)	Two elliptical Sersic profiles with initially unknown central position, peak surface brightness, scale radius, slope (Sersic index) and angle of orientation. One of these models the lens galaxy's core and the other models the fainter outer regions.\\
(iii)	A source, which is lensed by a PIEP+$\gamma$ lens with unknown parameters and blurred by the $5\times5$ pixel core of the TinyTim PSF (Figure~\ref{tinytim}). In Section~\ref{simple}, the source is modelled in a simple way as a sum of six elliptical Sersic profiles (in order to find a good initial value for the lens parameters), and in section~\ref{pixellated} the source is modelled as a pixellated grid with a prior distribution favouring non-negative pixel values and a dark background.\\
(iv)	Noise, to which we assign a Gaussian prior probability distribution with unknown standard deviation $\sqrt{\sigma_i^2 + \sigma_a^2 + \sigma_b^2 f_i}$ for each pixel, with $\sigma_a$ and $\sigma_b$ initially unknown and the ${\sigma_i}$ specified in advance to mask out the pixels with the most systematics.

One of the most difficult tasks in a source reconstruction problem is to find good values of the lens parameters to serve as the starting point for an MCMC run. If the source is pixellated, then exploration of the lens parameter space is slow because we effectively have to marginalize over thousands of source dimensions - so if we start with incorrect lens model values, the burn-in approach to the more plausible values will be slow. A good starting point for the lens can usually be obtained by running a much simpler version of the whole inference problem, for instance with a simpler model for the source, or by using only the quasar image positions and brightnesses as constraints. In the next section, we apply the latter approach to see how well the quasar images alone can constrain the lens model parameters. Later (Section~\ref{simple}), the extended images are taken into account by using a simply parameterised source model, where the number of source dimensions to be marginalized is only 36. Finally (Section~\ref{pixellated}), we use a pixellated model for the source in order to reconstruct its structure with minimal assumptions.
\section{How the quasar Images Constrain the Lens}
\subsection{Theory}
The four quasar images can constrain the lens model because we require that the four image positions lens back to the same point in the source plane. Actually, with an upper limit of $\sim$ 0.02 pc for the continuum source size \citep[e.g.][]{2005MNRAS.359..561W}, this exact requirement is too strong - we can really only insist that they lens back to within $\sim$ 3 microarcseconds of each other (using a concordance cosmology $\Omega_m = 0.27, \Omega_\Lambda = 0.73, H_0 = 71$ km s$^{-1}$Mpc$^{-1}$, \citet{2004ApJ...608...10N}). The results of this subsection are unchanged if we use a smaller quasar size of $10^{-3}$pc \citep{2008ApJ...676...80M} because the limiting factor is the accuracy of the astrometry, rather than the assumed quasar source size. The magnifications of the images can also provide some information (e.g. \citet{2002MNRAS.330L..15L} used the brightness of the third image of a quasar to argue for a model for the lens mass profile that creates a naked cusp in the source plane) although microlensing and absorption effects can limit the usefulness of including the magnifications as constraints for more typical systems.

The quasar positions were measured by fitting Gaussians to the peaks of the images. For the purposes of centroiding, this is an adequate approximation: the alternative is to calculate very high resolution simulated PSFs. Different (unknown) noise levels were assumed for each of the four images; however, the uncertainties in position were found to be $\sim$ 0.003 arc seconds for all of the four images. This corresponds to less than 1/10th of an image pixel.

To infer our PIEP+$\gamma$ parameters from the quasar positions, we first implemented an MCMC algorithm to explore the prior distribution for the lens parameters (Table~\ref{lenspriors}). Then we imposed a constraint on the probability distribution: that the expected value of the mean inter-pair distance, or spread, of the four points in the source plane upon back ray tracing, should be about $10^{-5}$ arc seconds. This modifies the probability distribution over the lens parameter space by a multiplicative factor $\exp\left(-k \times \texttt{Spread(Lens Parameters)}\right)$, where the value of $k$ is chosen so that the expectation value takes the value we wish to impose, $10^{-5}$ arc seconds. Conventionally, one would estimate the lens parameters by finding those values that minimise the spread in the source plane. Our probabilistic approach softens this constraint and implies that we sample from the range of lens models that reduce the scatter to about $10^{-5}$ arc seconds or less.

This approach assumes that we know the true exact image positions, although it can be extended to allow for uncertainty in the image positions, as follows. Denote the true unknown quasar image positions collectively by $X$, the estimated positions by $x$ and the lens parameter values by $L$. By the rules of probability theory, the posterior probability distribution for $L$ given $x$ (here, we assume that the known data are the $x$'s, rather than the entire image) can be written as:
\begin{eqnarray}
p(L|x) = \int p(L, X|x) dX \nonumber \\
= \int p(X|x) p(L|X,x) dX \nonumber \\
\end{eqnarray}
Since knowledge of $X$ would make $x$ irrelevant, this becomes
\begin{eqnarray}\label{qsopost}
p(L|x) = \int p(X|x) p(L|X) dX \nonumber \\
\propto \int p(X|x)p(L)\exp\left(-k \times \texttt{Spread}(X, L)\right) dX
\end{eqnarray}
Hence, the true image positions $\{X\}$ can be introduced as extra nuisance parameters to be estimated, and then we can sample the distribution under the integral sign in Equation~\ref{qsopost}. The small centroiding uncertainties of $\pm \sim$ 0.003 arcseconds were taken to specify the standard deviations of Gaussian probability distributions for $p(X|x)$.

The reader may wonder whether it would be more correct to introduce the unknown source plane position of the quasar as the nuisance parameters, calculate the image positions using the lens model, and use the $x$'s and their error bars to define a likelihood function. Whilst there is nothing wrong with this approach, it does involve the computationally challenging task of inverting the lens equations (Equation ~\ref{lenseqn}) to find the image positions. Our source plane spread approach is much easier to implement computationally, but relies on the unconventional step of directly assigning a posterior probability distribution: $p(X|x)$.

\subsection{Results}
To implement the inference described in the previous section, a Metropolis-Hastings sampler was written to target the posterior distribution of Equation~\ref{qsopost}. Unfortunately, this simple implementation had serious drawbacks. The joint posterior distribution for the lens parameters and true quasar image positions consists of long, thin, curved tunnels of high probability, and most standard sampling techniques have very poor mixing properties when sampling from such highly correlated distributions. To overcome this challenge, we used a different sampling technique, Linked Importance Sampling (LIS) \citep{lis}. LIS produces independent weighted samples from the target probability distribution. It only requires that we can independently sample from a simple distribution (e.g. the prior) and can define valid MCMC transitions with respect to distributions that are in some sense intermediate between the prior and posterior. For example, the common `annealing' approach of raising the likelihood to some power ($<$ 1) can be used within the LIS framework. The fact that each LIS run produces an {\it independent} sample from the target distribution makes it an attractive option for sampling from highly correlated distributions. The only possible pitfall is that the weights can vary significantly, such that the sample is completely dominated by only one or two points with large weight.

\begin{figure*}
\begin{center}
\includegraphics[scale=0.75]{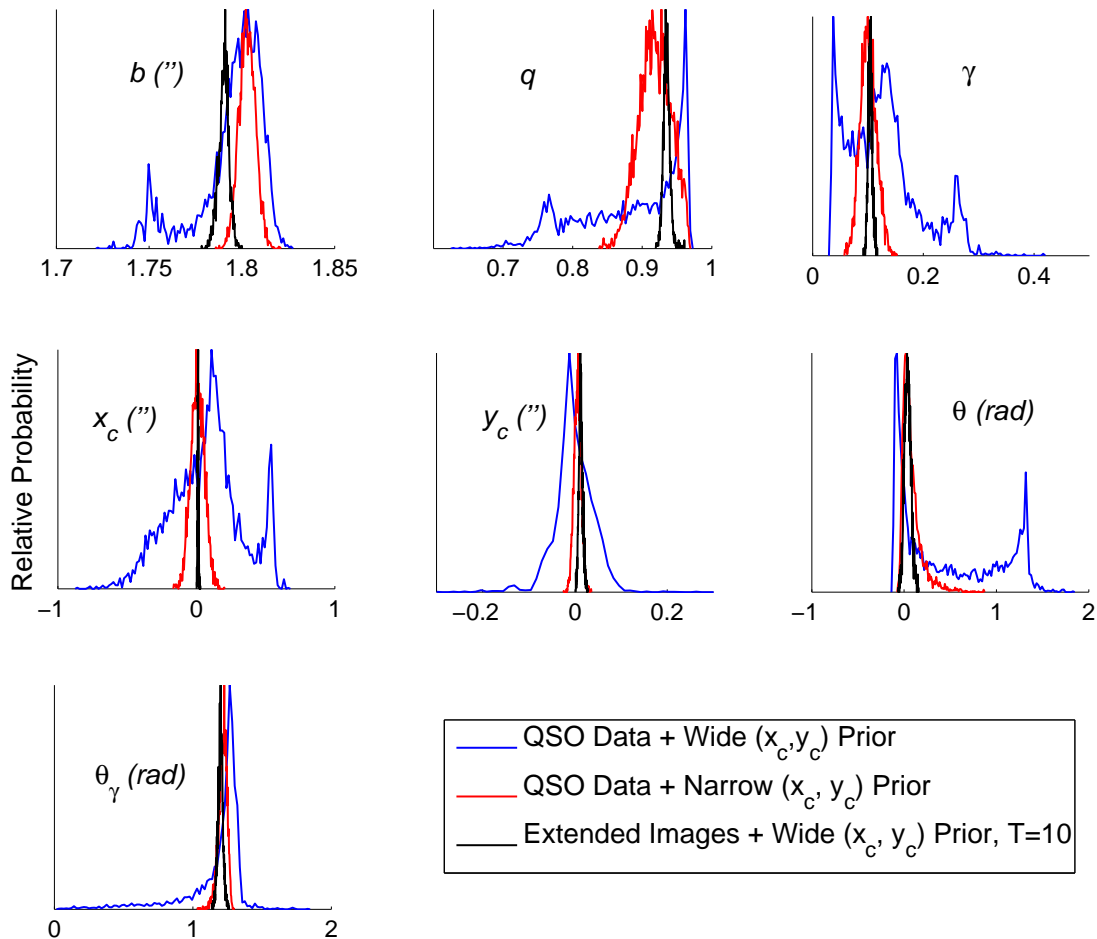}
\end{center}
\caption{Comparison of the inferred lens parameters (estimated marginal probability densities, unnormalized) based on three data sets: (1) The quasar image positions and uncertainties, with a weak prior distribution for the central position of the lens (blue curves). (2) The quasar image positions and uncertainties, with a strong prior ($\pm $ 1 pixel) on the central position of the lens (red curves), and (3) The entire image, and a pixellated source model, run at a temperature T=10 (black curves). Note that the parameter spaces for the two angles $\theta$ and $\theta_{\gamma}$ are periodic with period $\pi$.\label{comparison}}
\end{figure*}

We repeated this analysis twice: first, with the conservative priors on the central position of the lens - the priors in Table~\ref{lenspriors}. For comparison, we ran the algorithm with much more informative priors on the central position $(x_c,y_c)$ of the lens, an uncertainty of 1 pixel or $\pm 0.049$ arcseconds. The results are shown as the blue and red curves in Figure~\ref{comparison}. Far from uniquely determining the lens model, the quasar images have only managed to give moderately strong constraints on the overall strength of the lens ($b$) and the angle of orientation of the external shear ($\theta_{\gamma}$). For all other parameters, the marginal distributions are very wide, in some cases nearly as wide as the priors, so the quasars have provided only a small amount of information about them. In the seven dimensional space of the lens parameters, the quasar data yields a posterior distribution that contains long, narrow tunnels: while the volume of possible lens models is significantly decreased, the degeneracies inherent in lensing prevent precise inference about any single parameter.

With the stronger prior information about the central position (red curves), the marginal probability distributions tighten substantially, but not well enough to give a reliable lens parameter estimate for which we could trust a source reconstruction. In the next section, we use a simplistic model for the extended source in an attempt to get a good starting estimate for the lens model parameters. This estimate will then be used in the final run (Section~\ref{pixellated}) with a pixellated source plane.

\subsection{Simply-Parameterised Modelling of the Extended Source}\label{simple}
Rather than using pixels from the outset, we first modelled the source as the sum of six elliptical Gaussian-like (Sersic) profiles, with unknown brightness, orientation, central position, slope and ellipticity for each. Similarly, the lensing galaxy light profile was modelled as two elliptical Sersic profiles in the image plane. This simplified model reduces the dimensionality of the problem from thousands to 36, and makes it much more likely that a simple search algorithm can find something close to the optimal values for the lens model parameters. We used a simple Metropolis algorithm to derive estimates and uncertainties on the lens parameters. This estimate was used as a starting point for the chains in Section~\ref{pixellated}, where we use a pixellated source.

\begin{figure*}
\begin{center}
\includegraphics[scale=0.6]{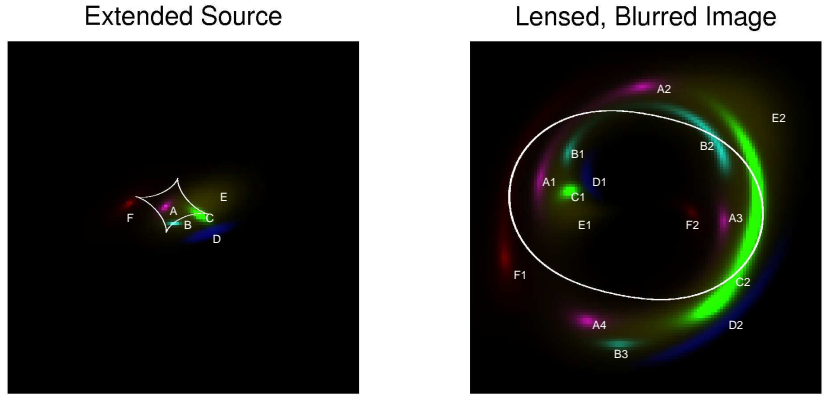}
\end{center}
\caption{Reconstruction of the extended source with six elliptical Gaussian-like Sersic profiles, shown on the same scale as the observed image. Most of the basic features of the observed image have been reproduced, but not their detailed shapes. The differences tell us where we should expect to see some additional substructure in the pixellated source model (see text). The positions of the caustics and critical curves are shown in white, although there should really be a small amount of uncertainty about their positions due to the uncertainty in the lens parameters.\label{mickey}}
\end{figure*}

A typical simple source model from the sample is shown in Figure~\ref{mickey}. Within this parameterisation, the uncertainties about the source are very small and Figure~\ref{mickey} can essentially be interpreted as the unique source reconstruction. The scales on the axes are the same for the source plane and the image plane, so an idea of the magnification can be obtained visually. Given this model, the data favour a highly complex source (since the blobs do not overlap to the extent that they become indistinguishable), lensed by a slightly elliptical lens whose centre is located close to the centre of the observed lensing galaxy. The individual Sersic profiles in the source shown in Figure~\ref{mickey} have been colour coded and labelled, making identification of the corresponding images easier. These labels will be used throughout the paper to refer to specific substructures in the source, and their corresponding images.

\begin{figure*}
\begin{center}
\includegraphics[scale=0.7]{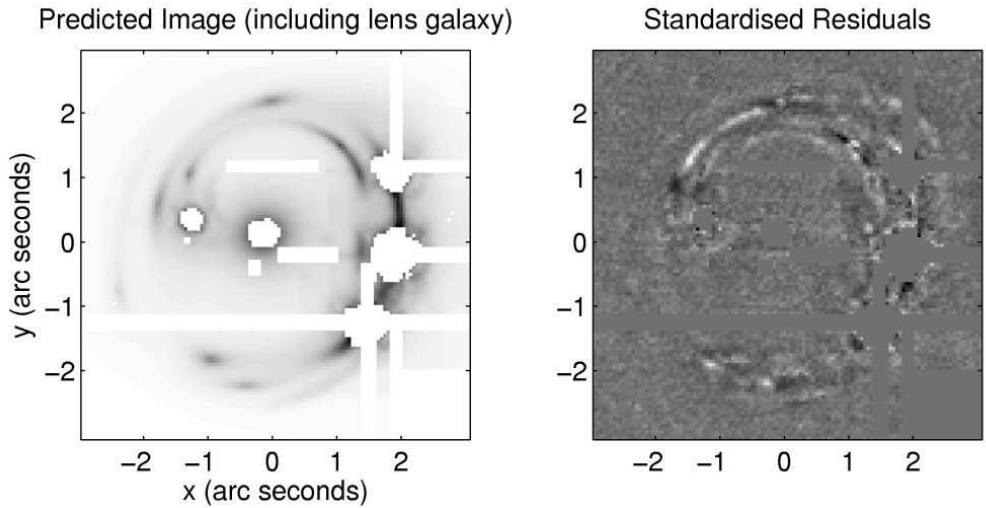}
\end{center}
\caption{Model-predicted lensed, blurred image, using a typical model sampled from the posterior distribution, and the simply-parameterised source model. On the right are the standardised residuals. While the basic features of the observed image are reproduced, there are details that have not been well modelled by the simply-parameterised source.\label{resid_simple}}
\end{figure*}

Comparing the image in Figure~\ref{mickey} to the observed one in the right hand panel of Figure~\ref{data}, we see that all of the basic structures observed have been reproduced by this simple model. However, the simply-parameterised model cannot reproduce the exact shapes of the observed images. For example, the part of the source labelled A should contain substructure, because we can see that the simple model has predicted a continuous image A1, yet the actual data contains blank gaps in some parts of that image. Another example is that image B2 has irregular brightness variations along its length, something the simply parameterised source model cannot reproduce. There are also some very faint additional structures that have been missed, such as a third faint inner ring below A2, that could be a continuation of the image D1. These differences can also be seen in the residuals in Figure~\ref{resid_simple}.

The bright ring (images C1 and C2) passes through the quasar image positions, so the quasar source is located inside source component C, but within the diamond caustic since the quasar has been imaged 4 times. However, due to uncertainty in the lens parameters, the quasar source cannot be located more accurately than this in the source plane. Source component C is moderately elliptical and is about 0.2 arcseconds in length, corresponding to a physical length scale of $\sim$ 1400 pc. The estimated magnifications for each component of the source are as follows: A (12.3), B (12.2), C (20.0), D (3.9), E (7.5), F (4.8), Quasar (45.0).

\section{Pixellated Source Model}\label{pixellated}
To obtain high resolution imaging of the source, we divided the source plane into a grid of 200$\times$200 pixels, each having a width of 0.015 arc seconds. The MCMC algorithm used is almost the same as that described in \citet{2006ApJ...651....8B}, and is described in greater detail there. Starting from a blank source, proposal changes are made that either add a bright ``atom'' of light to the source plane. Each atom has a position in the source plane indicating which pixel it is in, as well as a positive brightness, with exponential prior distribution. Small proposal transitions can be made which slightly adjust the parameters of an atom, in accordance with our chosen prior for these parameters. If the proposed source is a better fit, it is accepted, if it is a worse fit it is accepted with an acceptance probability equal to the ratio of the proposed likelihood to the current likelihood; this is just the standard Metropolis algorithm \citep{gregory} with a Massive Inference prior distribution \citep{massinf} for the source.

The prior expected value for an atom's flux can either be specified in advance, or incorporated as an additional hyperparameter to be estimated. The latter approach is attractive and has been used in the context of Gaussian priors for the source pixel values \citep{2006MNRAS.371..983S}. We chose the former approach for increased computational efficiency, and found that different values for this hyperparameter did not significantly change the final reconstruction, provided that the value was not so low that the reconstructed sky was bright or so high that the model cannot detect the presence of structures that are obviously real.

A straightforward implementation of any MCMC method is highly inefficient for the problem we have just posed. This is because we need to be able to explore the marginal posterior distribution (marginalising over possibly thousands of source parameters) for the lens parameters with reasonable efficiency. If we just alternated between source updates and lens updates, the lens parameters would only change in extremely tiny steps: for example, typically by $10^{-6}$, as far as the data and the current source will allow, so the lens model would explore the marginal distribution for lens parameters very slowly. Unfortunately, LIS was found to be unfeasible (the weights varied by orders of magnitude) due to the massive number of source parameters used. Hence, there was no realistic option other than to fix the lens model at a reasonable value and run the MCMC for the source variables only. We fixed the lens at the best values obtained from the simply-parameterised extended source model, and then ran a slow ``annealing'' \citep{annealing} simulation to determine a good lens model, which was then fixed, and the source space was explored at an annealing temperature of 1, to sample from the posterior distribution for the source pixel values given the lens. A large number of simulations with slightly different lens parameter values was done to verify that the source reconstruction is insensitive to slight uncertainties in the lens parameters. The results are presented in the next section.
\section{Results}
\subsection{Source}
\begin{figure*}
\begin{center}
\includegraphics[scale=0.7]{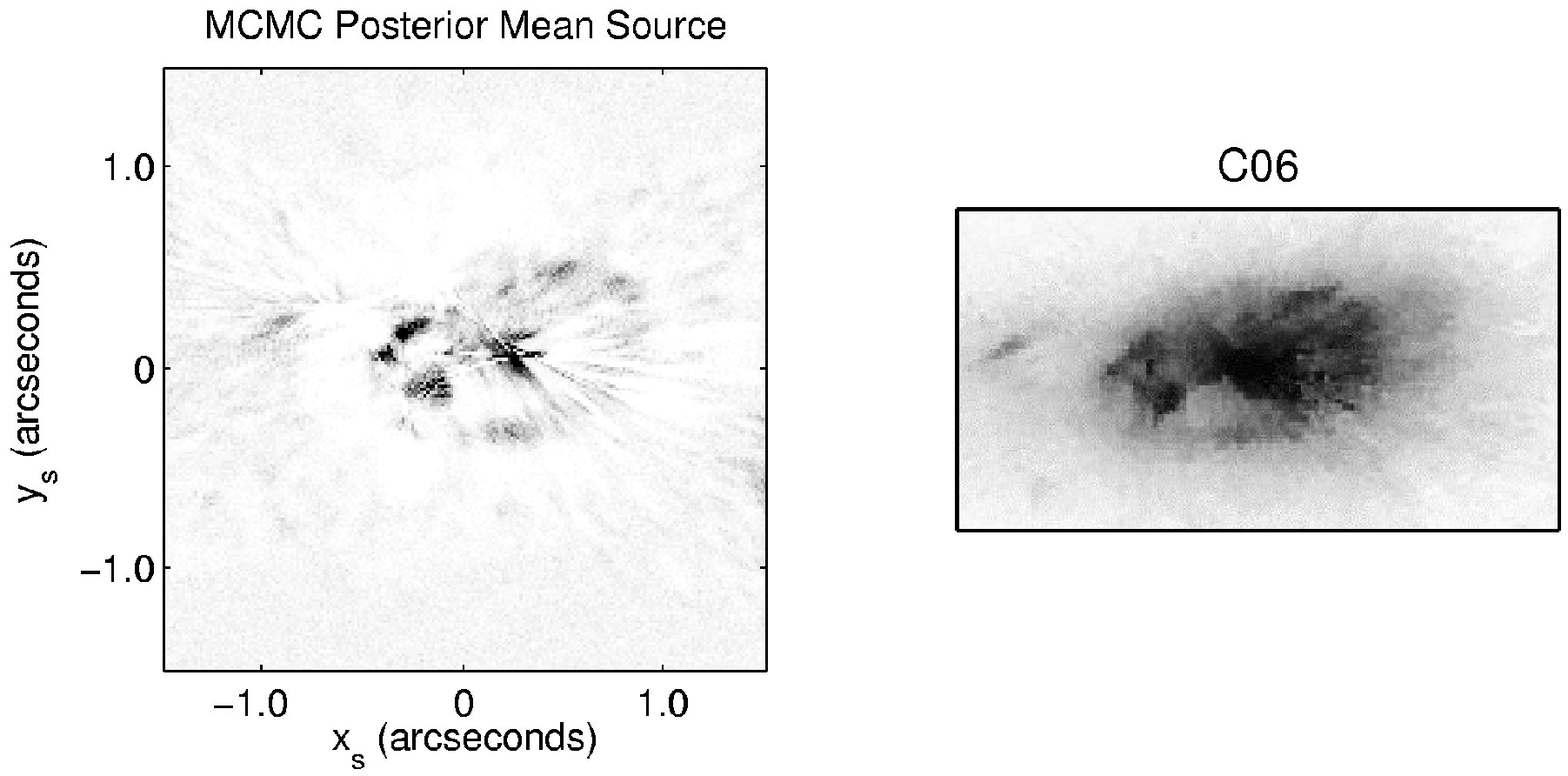}
\end{center}
\caption{Posterior mean source, with fixed lens parameters. On the right is a greyscale version of the original reconstruction by C06, for comparison. Note that the C06 reconstruction used data in three filters, which partially accounts for its more diffuse appearance: the compact substructures are most notable in the 555nm image (the subject of this paper).\label{results1}}
\end{figure*}

A final estimate for the source is shown in Figure~\ref{results1}, which is obtained by averaging all sources encountered by the Markov chain. Although the diversity of the samples encountered is an accurate representation of the level of uncertainty about any pixel values, it is inconvenient to present a large sample of images. Additionally, the spiky fluctuations caused by the Massive Inference prior remain in the sampled sources. Taking the mean of all of the sources provides a single estimate of the source profile that is optimal in the sense of minimising the expected square error; it also creates a more visually appealing smooth reconstruction where all of the spiky fluctuations in the individual samples have been averaged out. It is also quite natural for uncertain areas to be blurred in images, and for people viewing images to interpret a smooth image as possibly being caused by an underlying complex image.
\begin{figure*}
\begin{center}
\includegraphics[scale=0.7]{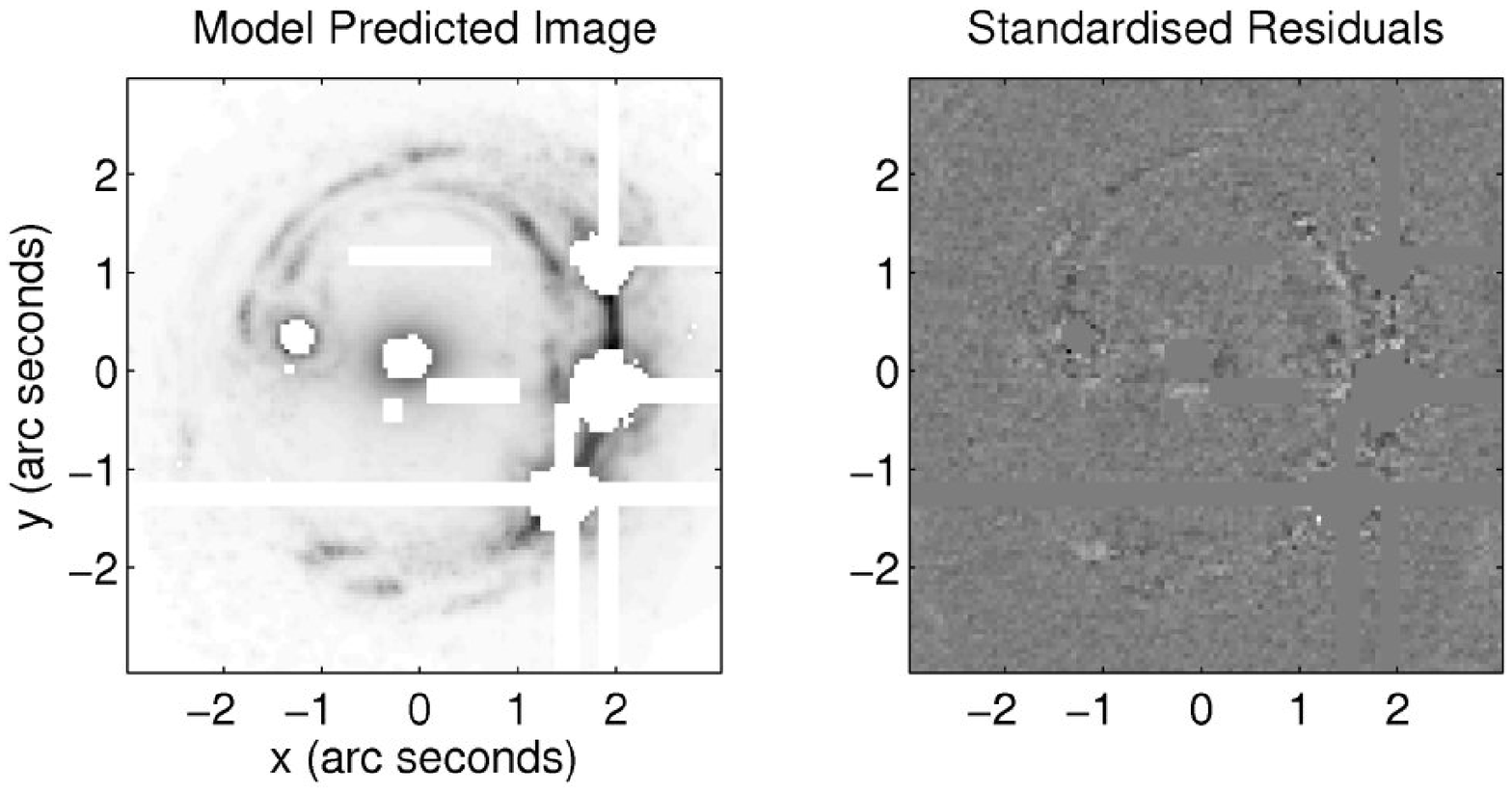}
\end{center}
\caption{Model-predicted lensed, blurred image, using a typical model sampled from the posterior distribution, and the pixellated source model. On the right are the standardised residuals.\label{resid}}
\end{figure*}

The positions of the major bright substructures in the source (Figure~\ref{results1}) are in agreement with those found by C06. However, our reconstruction presents a clearer view of the compact central source C, where the quasar resides. In the simply parameterised case, source C was found be elliptical. In the pixellated reconstruction, this part of the source has an elliptical component but with extra nearby structures that follow the caustic. The images of this extra structure lie within images C1 and C2. This suggests that the elliptical component is sufficient to explain the position of the ring C2 but not its brightness; the algorithm tries to account for this by adding extra source flux in regions that have images within C2 but only within those parts of C1 that have been masked out. It seems more plausible that image C1 is more affected by dust absorption than C2, rather than the source coincidentally following the caustic. In principle, we could parameterise the unknown dust profile of the lens galaxy and simultaneously estimate this from the data (as done by e.g. \citet{2008arXiv0804.2827S}), but this is beyond the scope of the present paper.

The predicted gap in the middle of source component A is also clearly present in our reconstruction, whereas it is much less clear, if present at all, in the reconstruction by C06. Source component E also contains some substructure of its own. In addition, the surface brightness contrast between these features and the diffuse background is much greater in our reconstruction, although this is simply because we are focusing on the 555nm image. At other wavelengths, the compact substructures are less pronounced.

\begin{figure*}
\begin{center}
\includegraphics[scale=0.6]{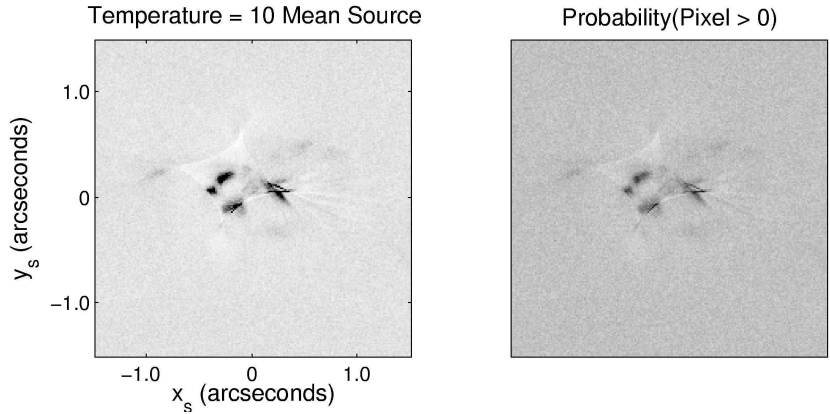}
\end{center}
\caption{Mean source encountered by a chain run at a temperature of 10. This allows the chain to explore the parameter space more freely, reducing the effect of unmodelled systematics present in the Temperature=1 reconstruction (albeit by discarding information). Most of the substructures are still sharply resolved. On the right, the posterior probability that any pixel is nonzero is plotted. White corresponds to 0, while black corresponds to 1. While it is very difficult to be confident about any particular pixel (due to the finite information content of the image), groups of pixels with consistently high values of this probability represent a secure detection.\label{temp10}}
\end{figure*}

To reduce these systematic effects, we repeated the simulations at an annealing temperature of 10, to allow freer exploration of the parameter space. This is an ad-hoc device that is not as justified as explicit modelling of dust \citep{2008arXiv0804.2827S}, but is helpful nonetheless. At a temperature of 10, it also becomes computationally feasible to free the lens parameters. The results, shown in Figure~\ref{temp10}, still show clear images of the bright, compact substructures that are present in the source. These bright substructures account for all of the bright images that are visible in the data; the diffuse emission that the T=10 analysis misses is caused by a thicker, fainter ring (image E2) that is not the most striking feature of the image.

The reconstruction produced by the high temperature run still contains source plane structure that follows the caustic. This suggests that increasing the temperature has not completely negated the effect of dust. However, the raised temperature simulation still results in model images that reproduce the positions and shapes of all of the details in the observed Einstein ring, while being more permissive about their fluxes. Thus, the source reconstruction in Figure~\ref{temp10} is a good model for the positions and (excluding the central source C) shapes of the source galaxy substructures.

This system is probably one of the most complex Einstein Rings known, with a spectacular number of distinct extended images. Hence, it is unsurprising that the source galaxy morphology is also very complex. The redshift of the source ($z=0.658$) implies that this light was emitted in the near UV (wavelength of 335 nm), suggesting that these structures are mapping the star forming regions of the source galaxy. With the assumed cosmology, the length scale in the source plane is $\sim$ 6.959 kpc/arcsecond. The bulk of the galaxy is just over 1 arc second across, so the entire source is about 8 kpc across; the source is a medium-sized galaxy. Compared to C06, our reconstruction is devoid of a large amount of extended emission. This is probably caused by some light from the lens galaxy being attributed to the source in their analysis. The surface brightness ratio of the compact structures to the diffuse emission can be estimated visually by simply looking at the image; where the contrast is a lot stronger than it appears to be in the C06 reconstruction. There is also evidence for a companion dwarf galaxy 2.4 kpc away and roughly 700 pc in diameter, to the left of the main galaxy (source F). We have also found that the quasar resides within an elliptical ($q\sim 0.5$) region of young stars that is about 1400 pc in extent.

\subsection{Lens Parameters}
When running the MCMC simulation at a temperature of 10, inferring the lens parameter uncertainties becomes computationally feasible. Thus, we can compare the lens constraints derived from the extended images with those derived from the point-like images. The marginal posterior distribution for the lens parameters, with a pixellated source, are shown as the black curves in Figure~\ref{comparison}. This shows clearly that the extended images provide significantly stronger constraints on the lens model parameters in this system, in contrast to the claim made by C06 that the opposite is true, and in agreement with \citet{2001ApJ...547...50K}. Note that the distributions for $b$ inferred from the quasar astrometry and the extended image reconstruction overlap only slightly. This is because the very small quoted astrometric uncertainties of 0.003'' (less than 1/10th of an image pixel) do not take into account known systematic effects such as the presence of background flux from the lens galaxy and the Einstein Ring, the fact that the peak of the PSF is not really a Gaussian, and the fact that the lens is not {\it really} a PIEP+$\gamma$. This small disagreement means that the parameters inferred from the extended images are only marginally capable of giving quasar image positions correctly to within 0.003''. They are, however, capable of reproducing the positions to within a relaxed tolerance of $\sim$ 1/3 of a pixel. The axis ratio $q$, describing the ellipticity of the lens potential, is found to be 0.935 $\pm$ 0.005. Clearly, the data rule out $q=1$ and therefore Singular Isothermal Sphere + $\gamma$ models (this is still the case even when only the quasar data is used). If we had assigned a delta function of prior probability at $q=1$, the data would downgrade its posterior probability significantly when compared to any realistic diffuse prior for $q$.

The mass of the lens can be calculated from these results. For the purposes of this calculation, the lens is approximated as circular. For an isothermal sphere, the deflection angles at the Einstein radius of the ring (now equal to $b$) is simply $b$. For a point mass at the origin, the required mass to produce the same deflection would be $b^2$. Since lensing obeys Gauss' law, $b^2$ must also be the amount of mass enclosed within the ring. An alternative approach would be to calculate the nondimensional density, which proportional to the 2-D Laplacian of the lens potential \citep{book}. In scaled units, the total mass of the lens is therefore $b^2=3.205 \pm 0.01$, where the uncertainty was found, as usual, by considering an ensemble of lens models from the MCMC output and calculating the mass for each. However, systematics introduced by the approximations in computing this value will probably overwhelm this quoted uncertainty. The mass unit that would give an Einstein Radius of 1 arc second needs to be computed to convert this figure into physical units. When this is done, the estimated mass of the lensing galaxy (within the ring; the total mass outside of this is very poorly contrained by lensing) is found to be $(6.95 \pm 0.02)\times 10^{11}$ $M_0$. With the isothermal assumption, the velocity dispersion is $\sim$ 350 km s$^{-1}$. From the lens galaxy light profile parameters, we find that the flux of the lens galaxy (within the Einstein Ring) is close to 50\% of the flux of the brightest QSO image, which has a magnitude of 17.74 in the 555nm filter. At a redshift of $z$=0.295, the luminosity distance to the lens (with our assumed cosmology) is 1510 Mpc. Thus, the average (within the ring) mass to light ratio of the lens galaxy is found to be 8.8 $M_0/L_0$.

The lens potential is only slightly elliptical and is centred near the centre of the lens galaxy, if the centre is defined by the brightest pixel of the galaxy core. All of these conclusions are similar to those made of ER 0047-2808, and are probably typical features of Einstein rings and all other systems with single galaxy lenses. This lends more support to the often used assumption that the centre of a lens galaxies light profile is also the point at which the lens model should be centred. It remains unclear how the galaxy light profile information could be taken into account in a more complex kiloparametric model of the lens; that is a topic for further research. The lens light profile parameters are presented in Table~\ref{lenslight}.

\begin{table*}
\begin{center}
\caption{Lens galaxy light profile parameters, and the noise parameters. When the dimensions are those of surface brightness, the units are 1/20th of the flux of the brightest QSO image (mag 17.74) per square arcsecond. Length units are arcseconds.}\label{lenslight}
\begin{tabular}{lcc}
\hline Parameter & Component 1 (compact) & Component 2 (diffuse)\\
\hline
Peak Brightness & 2.99 $\pm$ 0.31 & 0.283 $\pm$ 0.007\\
Ellipticity & 0.85 $\pm$ 0.01 & 0.80 $\pm$ 0.02\\
x Central Position & -0.166 $\pm$ 0.004 & 0.45 $\pm$ 0.03\\
y Central Position &  0.156 $\pm$ 0.003 & 0.07 $\pm$ 0.01\\
Scale Radius & 0.33 $\pm$ 0.03 & 2.65 $\pm$ 0.03\\
Angle of Orientation & 3.6 $\pm$ 2.4$^\circ$ & -16 $\pm$ 2$^\circ$\\
Slope $\alpha$ & 1.3 $\pm$ 0.1 & 2.70 $\pm$ 0.14\\
$\sigma_a$ & 0.0479 $\pm$ 0.0007\\
$\sigma_b$ & 0.009 $\pm$ 0.002 \\
\end{tabular}
\medskip\\
\end{center}
\end{table*}

A preliminary investigation of the time delays predicted by our lens model suggests that it does not exactly reproduce the time delays measured by \citet{timedelay}. Since this system exhibits significant microlensing \citep{2008arXiv0805.4492C, 2008RMxAC..32...83S}, the time delay measurements are uncertain, but it is possible that the PIEP+$\gamma$ lens model will be ruled out by further observations. This would not be catastrophic for the present study, for several reasons. Firstly, source reconstructions tend to be insensitive to slight misspecification of the lens model \citep[e.g.][]{wayth}. Secondly, all parameterisations are false. We already know from prior information that the lens is not {\it really} a PIEP+$\gamma$ model. All modelling can only consider a single ``slice'' through a full hypothesis space, and the conclusions reached on that slice may or may not be representative of the full space. They often are, but there are never any guarantees.

\section{Conclusions and Further Work}
In this paper, we have presented a detailed gravitational lens reconstruction of the optical extended source in the Einstein Ring RXS J1131-1231. The source is a medium sized galaxy ($\sim$ 8 kpc in visible extent) with several compact bright emission regions. The substructures we found are in general agreement with those found by C06 in terms of their position, but we have shown that they are brighter and more compact than was previously determined. In addition, our reconstruction provides a clearer view of the substructures, including near the central regions of the source. The quasar resides in a bright emission region with an extent of about $\sim$ 0.15 arcseconds or 1 kpc. It should be noted that the wavelength of the observations in the rest frame is 335 nm, so this reconstruction traces regions of recent star formation in the source galaxy.

We have also directly compared point images vs extended images with regard to how well each is able to constrain the lens model. We found that there is a significant gain to be made in taking into account all of the information from the extended images. It has been suggested that this is not true in general \citep{2007arXiv0710.3159F}, although it really depends on the resolution and number of extended images, which in this case is high. Certainly, in using both, there is nothing to lose but CPU cycles. This system has the potential to become one of the most well-constrained gravitational lenses, with multiple images of the extended ring, quasar image positions and flux ratios in multiple bands, and time delay measurements available \citep{timedelay, keeton}. Hence, it should be possible to carry out a detailed kiloarametric study of its mass profile to shed some light on the dark matter halo of the lens galaxy.

This paper was based on a single image of this system, the 555nm ACS image. Other HST images at different wavelengths (814nm, 1.6$\mu$m) are available (C06) and can further constrain the lens model. Simultaneous multi-wavelength reconstructions are now becoming routine \citep[e.g.][]{2007ApJ...671.1196M}. However, all of the structures in these images are in the same locations, and so a multi-wavelength reconstruction would not produce significantly different conclusions to those reached here. C06 note that in the near infrared image, the compact bright images are less pronounced compared to the diffuse background, which is what would be expected if the substructures are regions dense in hot young stars.

This study has relied on a number of common assumptions that future research will seek to relax. Extending lens reconstruction techniques to incorporate kiloparametric models of the source and the lens simultaneously is an ambitious task, but some steps are already being taken in that direction \citep{2008arXiv0804.2827S}. Flexible lens modelling plus information from time delay measurements and other sources would be extremely valuable for studies of galaxy dark matter haloes. Also, explicit modelling of dust absorption by the lens galaxy is proving to be an important ingredient in the inversion of Einstein Rings and would be an essential part of future work on this system.

\section*{Acknowledgments}
BJB thanks Olivia Ross for encouragement, and Dennis Stello for allowing me to use his fast computer. This research is supported under the Australian Research Council's Discovery funding scheme (project number DP0665574) and is part of the Commonwealth Cosmology Initiative (CCI). The authors would like to thank Jean-Francois Claeskens and Dominique Sluse for providing the ACS data. The constructive comments of the anonymous referee helped us to improve the paper significantly.
%\label{lastpage}


\begin{thebibliography}{99}
\bibitem[\protect\citeauthoryear{Blandford 
\& Kochanek}{1987}]{piepref} Blandford R.~D., Kochanek C.~S., 1987, ApJ, 321, 658 

\bibitem[\protect\citeauthoryear{Brewer \& 
Lewis}{2006b}]{2006ApJ...651....8B} Brewer B.~J., Lewis G.~F., 2006, ApJ, 
651, 8 

\bibitem[\protect\citeauthoryear{Brewer \& 
Lewis}{2006a}]{2006ApJ...637..608B} Brewer B.~J., Lewis G.~F., 2006, ApJ, 
637, 608 

\bibitem[\protect\citeauthoryear{Chartas et 
al.}{2008}]{2008arXiv0805.4492C} Chartas G., Kochanek C.~S., Dai X., 
Poindexter S., Garmire G., 2008, arXiv, 805, arXiv:0805.4492 

\bibitem[\protect\citeauthoryear{Claeskens et 
al.}{2006}]{2006A&A...451..865C} Claeskens J.-F., Sluse D., Riaud P., 
Surdej J., 2006, A\&A, 451, 865 

\bibitem[\protect\citeauthoryear{Dye et al.}{2008}]{2008arXiv0804.4002D} 
Dye S., Evans N.~W., Belokurov V., Warren S.~J., Hewett P., 2008, arXiv, 
804, arXiv:0804.4002 

\bibitem[\protect\citeauthoryear{Dye \& Warren}{2005}]{2005ApJ...623...31D} 
Dye S., Warren S.~J., 2005, ApJ, 623, 31

\bibitem[\protect\citeauthoryear{Ferreras, Saha, \& 
Burles}{2007}]{2007arXiv0710.3159F} Ferreras I., Saha P., Burles S., 2007, 
arXiv, 710, arXiv:0710.3159

\bibitem[\protect\citeauthoryear{Kirkpatrick, Gelatt, \& 
Vecchi}{1983}]{annealing} Kirkpatrick S., Gelatt C.~D., Vecchi 
M.~P., 1983, Sci, 220, 671 

\bibitem[\protect\citeauthoryear{Gregory}{2005}]{gregory} 
Gregory P.~C., 2005, Bayesian Logical Data Analysis for the Physical Sciences, Cambridge University Press

\bibitem[\protect\citeauthoryear{Keeton \& Moustakas}{2005}]{keeton} 
Keeton, C.~R. and Moustakas, L.~A. arXiv:0805.0309

\bibitem[\protect\citeauthoryear{Kochanek, Keeton, 
\& McLeod}{2001}]{2001ApJ...547...50K} Kochanek C.~S., Keeton C.~R., McLeod B.~A., 2001, ApJ, 547, 50 

\bibitem[\protect\citeauthoryear{Koopmans}{2005}]{2005MNRAS.363.1136K} 
Koopmans L.~V.~E., 2005, MNRAS, 363, 1136 

\bibitem[\protect\citeauthoryear{Koopmans et 
al.}{2006}]{2006ApJ...649..599K} Koopmans L.~V.~E., Treu T., Bolton A.~S., 
Burles S., Moustakas L.~A., 2006, ApJ, 649, 599 

\bibitem[\protect\citeauthoryear{Krist}{1995}]{tinytimreference} Krist 
J., 1995, ASPC, 77, 349 

\bibitem[\protect\citeauthoryear{Lewis et al.}{2002}]{2002MNRAS.330L..15L} 
Lewis G.~F., Carilli C., Papadopoulos P., Ivison R.~J., 2002, MNRAS, 330, 
L15 

\bibitem[\protect\citeauthoryear{Marshall et 
al.}{2007}]{2007ApJ...671.1196M} Marshall P.~J., et al., 2007, ApJ, 671, 
1196 

\bibitem[\protect\citeauthoryear{Morgan et al.}{2006}]{timedelay} 
Morgan N.~D., Kochanek C.~S., Falco E.~E., Dai X., 2006, astro, 
arXiv:astro-ph/0605321 

\bibitem[\protect\citeauthoryear{Morgan et al.}{2008}]{2008ApJ...676...80M} 
Morgan C.~W., Eyler M.~E., Kochanek C.~S., Morgan N.~D., Falco E.~E., 
Vuissoz C., Courbin F., Meylan G., 2008, ApJ, 676, 80 

\bibitem[\protect\citeauthoryear{Neal}{2005}]{lis} Neal, R.~M., 2005, Estimating Ratios of Normalizing Constants Using Linked Importance Sampling, arXiv:math/0511216v1

\bibitem[\protect\citeauthoryear{Nolta et al.}{2004}]{2004ApJ...608...10N} 
Nolta M.~R., et al., 2004, ApJ, 608, 10 

\bibitem[\protect\citeauthoryear{Pettini et 
al.}{2000}]{2000ApJ...528...96P} Pettini M., Steidel C.~C., Adelberger 
K.~L., Dickinson M., Giavalisco M., 2000, ApJ, 528, 96 

\bibitem[\protect\citeauthoryear{Schneider, Ehlers, \& 
Falco}{1992}]{book} Schneider P., Ehlers J., Falco E.~E., 
1992, Gravitational Lenses

\bibitem[\protect\citeauthoryear{Skilling}{1998}]{massinf} 
Skilling J., 1998, Massive Inference and Maximum Entropy, in Maximum Entropy and Bayesian Methods, Kluwer Academic Publishers, Dordrecht/Boston/London

\bibitem[\protect\citeauthoryear{Sluse et al.}{2003}]{2003A&A...406L..43S} 
Sluse D., et al., 2003, A\&A, 406, L43 

\bibitem[\protect\citeauthoryear{Sluse et al.}{2008}]{2008RMxAC..32...83S} 
Sluse D., Claeskens J.-F., Hutsem{\'e}kers D., Surdej J., 2008, RMxAC, 32, 
83 


\bibitem[\protect\citeauthoryear{Sugai et al.}{2007}]{2007ApJ...660.1016S} 
Sugai H., Kawai A., Shimono A., Hattori T., Kosugi G., Kashikawa N., Inoue 
K.~T., Chiba M., 2007, ApJ, 660, 1016 

\bibitem[\protect\citeauthoryear{Suyu et al.}{2006}]{2006MNRAS.371..983S} 
Suyu S.~H., Marshall P.~J., Hobson M.~P., Blandford R.~D., 2006, MNRAS, 
371, 983 

\bibitem[\protect\citeauthoryear{Suyu et al.}{2008}]{2008arXiv0804.2827S} 
Suyu S.~H., Marshall P.~J., Blandford R.~D., Fassnacht C.~D., Koopmans 
L.~V.~E., McKean J.~P., Treu T., 2008, arXiv, 804, arXiv:0804.2827 

\bibitem[\protect\citeauthoryear{Treu et al.}{2008}]{2008arXiv0806.1056T} 
Treu T., Gavazzi R., Gorecki A., Marshall P.~J., Koopmans L.~V.~E., Bolton 
A.~S., Moustakas L.~A., Burles S., 2008, arXiv, 806, arXiv:0806.1056 

\bibitem[\protect\citeauthoryear{Warren 
\& Dye}{2003}]{2003ApJ...590..673W} Warren S.~J., Dye S., 2003, ApJ, 590, 673 

\bibitem[\protect\citeauthoryear{Wayth et al.}{2005}]{wayth} 
Wayth R.~B., Warren S.~J., Lewis G.~F., Hewett P.~C., 2005, MNRAS, 360, 
1333 

\bibitem[\protect\citeauthoryear{Wayth, O'Dowd, 
\& Webster}{2005}]{2005MNRAS.359..561W} Wayth R.~B., O'Dowd M., Webster R.~L., 2005, MNRAS, 359, 561 

\end{thebibliography}
\end{document}